\documentclass[traditabstract]{aa} %
\usepackage{longtable,lscape}
\usepackage{natbib}
\usepackage{graphicx}
\def\ms{\hbox{\,m\,s$^{-1}$}}         
\def\m2s2{\hbox{\,m$^{2}$\,s$^{-2}$}} 
\def\kms{\hbox{\,km\,s$^{-1}$}}       
\def\Msun{\hbox{$\mathrm{M}_{\odot}$}}             
\def\Rsun{\hbox{$\mathrm{R}_{\odot}$}}
\def\Mjup{\hbox{$\mathrm{M}_{\rm Jup}$}}

\begin{document}
\title{The GAPS programme with HARPS-N@TNG \\
          IV: A planetary system around XO-2S
          \thanks{Based on observations made with the Italian Telescopio
                  Nazionale Galileo (TNG) operated on the island of La Palma
                  by the Fundacion Galileo Galilei of the INAF
                  at the Spanish
                  Observatorio del Roque de los Muchachos of the IAC
                  in the frame of the program
                  Global Architecture of Planetary Systems (GAPS),
                  and on observations made at Asiago, Serra La Nave, and
                  Valle D'Aosta Observatories.
                  }}

\author{
S. Desidera    \inst{1},
A.S. Bonomo    \inst{2},
R.U. Claudi    \inst{1},
M. Damasso     \inst{2,3},
K. Biazzo      \inst{4},
A. Sozzetti    \inst{2},
F. Marzari     \inst{5,1},
S. Benatti     \inst{1},
D. Gandolfi    \inst{4,6},
R. Gratton     \inst{1},
A.F. Lanza     \inst{4}
V. Nascimbeni  \inst{7,1},
G. Andreuzzi   \inst{8},
L. Affer       \inst{9},
M. Barbieri    \inst{7},
L.R. Bedin     \inst{1},
A. Bignamini   \inst{10},
M. Bonavita    \inst{1},
F. Borsa       \inst{11},
P. Calcidese   \inst{3},
J.M. Christille \inst{3,12},
R. Cosentino   \inst{4,8},
E. Covino      \inst{13},
M. Esposito    \inst{14},
P. Giacobbe    \inst{2},
A. Harutyunyan \inst{8},
D. Latham      \inst{15},
M. Lattanzi    \inst{2},
G. Leto        \inst{4},
G. Lodato      \inst{16},
C. Lovis       \inst{17},
A. Maggio      \inst{9},
L. Malavolta   \inst{7,17},
L. Mancini     \inst{18},
A.F. Martinez Fiorenzano \inst{8},
G. Micela      \inst{9},
E. Molinari    \inst{8,19},
C. Mordasini   \inst{18},
U. Munari      \inst{1},
I. Pagano      \inst{4},
M. Pedani      \inst{8},
F. Pepe        \inst{17},
G. Piotto      \inst{7,1},
E. Poretti     \inst{11},
M. Rainer      \inst{11},
I. Ribas       \inst{20},
N.C. Santos    \inst{21,22},
G. Scandariato \inst{4},
R. Silvotti    \inst{2},
J. Southworth  \inst{23},
R. Zanmar Sanchez      \inst{4}
           }

\institute{INAF -- Osservatorio Astronomico di Padova,  Vicolo dell'Osservatorio 5, I-35122, Padova, Italy
\and INAF -- Osservatorio Astrofisico di Torino, Via Osservatorio 20, I-10025, Pino Torinese, Italy
\and Osservatorio Astronomico della Regione Autonoma Valle d'Aosta,  Fraz. Lignan 39, I-11020, Nus (Aosta), Italy
\and INAF -- Osservatorio Astrofisico di Catania, Via S.Sofia 78, I-95123, Catania, Italy
\and Dipartimento di Fisica e Astronomia Galileo Galilei -- Universit\`a di Padova, Via Marzolo 8, I-35131 Padova, Italy
\and Landessternwarte K\"onigstuhl, ZAH, Universit\"at Heidelberg, K\"onigstuhl 12, D-69117 Heidelberg, Germany
\and Dip. di Fisica e Astronomia Galileo Galilei -- Universit\`a di Padova, Vicolo dell'Osservatorio 2, I-35122, Padova, Italy
\and Fundaci\'on Galileo Galilei - INAF, Rambla Jos\'e Ana Fernandez P\'erez 7, E-38712 Bre\~na Baja, TF - Spain
\and INAF -- Osservatorio Astronomico di Palermo, Piazza del Parlamento, Italy 1, I-90134, Palermo, Italy
\and INAF -- Osservatorio Astronomico di Trieste, via Tiepolo 11, I-34143 Trieste, Italy
\and INAF -- Osservatorio Astronomico di Brera, Via E. Bianchi 46, I-23807 Merate (LC), Italy
\and Dept. of Physics, University of Perugia, via A. Pascoli 1, 06123, Perugia, Italy
\and INAF -- Osservatorio Astronomico di Capodimonte, Salita Moiariello 16, I-80131, Napoli, Italy
\and Instituto de Astrofisica de Canarias, C/Via Lactea S/N, E-38200 La Laguna, Tenerife, Spain
\and Harvard-Smithsonian Center for Astrophysics, 60 Garden Street, Cambridge, MA 02138
\and Dipartimento di Fisica, Universit\`a  di Milano, Via Celoria 16, I-20133 Milano, Italy
\and Obs. Astronomique de l'Univ. de Geneve, 51 ch. des Maillettes -  Sauverny, CH-1290, Versoix, Switzerland
\and Max-Planck-Institut f\"ur Astronomie, K\"onigstuhl 17, D-69117, Heidelberg, Germany
\and INAF - IASF Milano, via Bassini 15, I-20133 Milano, Italy
\and Inst. de Ciencies de l'Espai (CSIC-IEEC), Campus UAB, Facultat de Ciencies, 08193 Bellaterra, Spain
\and Centro de Astrof{\'\i}sica, Universidade do Porto, Rua das Estrelas,  4150-762 Porto, Portugal
\and Departamento de F{\'\i}sica e Astronomia, Faculdade de Ci\^encias,  Universidade do Porto, Portugal
\and Astrophysics Group, Keele University, Staffordshire, ST5 5BG, UK
              }

\date{}

\abstract{We performed an intensive radial velocity monitoring of XO-2S, the wide companion of the transiting
planet-host XO-2N, using HARPS-N at TNG in the framework of the GAPS programme.
The radial velocity measurements indicate
the presence of a new planetary system formed by a planet that is slightly more massive than Jupiter
at 0.48 au and a Saturn-mass planet at 0.13 au. Both planetary orbits are moderately eccentric and were
found to be dynamically stable. There are also indications of a long-term trend in the radial velocities.
This is the first confirmed case of a wide binary whose components both host planets, 
one of which is transiting, which makes the XO-2 system a unique laboratory for understanding 
the diversity of planetary systems.}


\authorrunning{S. Desidera et al.}
\titlerunning{GAPS IV: A planetary system around XO-2S}

\maketitle

\section{Introduction}
\label{s:intro}

The discoveries of extrasolar planets in the past two decades have revealed a surprising variety of system architectures and
planet characteristics. The wide diversity of the outcomes of the planet formation process is linked
to the properties of the host stars, the characteristics of the circumstellar disks, and
the effects of the environment in which the stars and their planets form and evolve \citep{2012A&A...541A..97M}.

A better understanding of the stochastic factors that affect planet formation
can be achieved by observing wide binary systems with similar components. This allows
some of the relevant variables (chemical composition, birth environment, age) to be the same for both
components. On the other hand, the presence of stellar companions can affect 
the dynamical evolution of planetary systems.
\citet{2013Natur.493..381K} showed that even very wide binaries ($a \sim 1000$~au) can have a significant impact on
the survival and orbital properties of any planets they might host, because of 
the continuous evolution of the binaries'
orbital elements caused by interactions with passing stars and Galactic tides.
Indeed, some differences in the properties of planets in binaries with respect to those orbiting single stars
have been identified, such as the excess of massive close-in planets \citep{2007A&A...462..345D}
and of planets on highly eccentric orbits \citep{2013Natur.493..381K}.

With a few exceptions \citep{2011A&A...533A..90D,2009PASJ...61...19T},
radial velocity (RV) surveys have historically excluded known binaries from their samples, or included only
one component of optically resolved systems, depending on the adopted selection criteria.
Furthermore, the probability to have planets transiting each
component of a binary system is low, and resolving the binary is challenging for stars at the typical
distance of the targets of transit surveys.
Nevertheless, some candidates have been identified by the {\it Kepler} mission, with  \object{Kepler-132}
being the most promising one \citep{2014ApJ...784...44L}.
However, the available data only allow the conclusion that the planet candidates orbit different components.
The identification of the host of each transiting object remains ambiguous, and further characterization
is hampered by the faintness of the stars and their small angular separation.

Another possibility is to search for planets around the companions of known planet hosts. Stars with planets are often
the subject of searches for stellar companions \citep[e.g.,][]{2006A&A...456.1165C,2014MNRAS.439.1063M},
or common proper motion stars may be known well before the discovery of planets \citep{2006ApJ...646..523R}.
In this letter we present the results of intensive RV monitoring of the K0 star \object{XO-2S} (TYC 3413-210-1),
performed using the spectrograph HARPS-N \citep{2012SPIE.8446E..1VC} at the Telescopio Nazionale Galileo (TNG)
as part of the programme Global Architecture of Planetary Systems \citep[GAPS,][]{2013A&A...554A..28C,2013A&A...554A..29D}.
XO-2S ($V$=11.12 mag; $B-V$=0.79) is the wide (30$^{\prime\prime}$, $\sim$4000~au projected separation) companion
of \object{XO-2N}, which was found to host a transiting planet of mass $0.5~M_{\rm J}$ and orbital period
2.5~d \citep{2007ApJ...671.2115B}. The two components are very similar ($\Delta R \sim 0.04$ mag), and both are super-metal-rich in composition.
Our observations allowed us to detect two giant planets around XO-2S, both at orbital separations larger than the hot
Jupiter around XO-2N, thus making the XO-2 system the first confirmed case of a 
wide binary whose components both host planets, one of which is transiting.
Here we report the RV measurements, stellar activity and line profile indicators,
ancillary photometric observations taken at the Asiago, Valle d'Aosta and Serra La Nave
Observatories supporting the Keplerian origin of the RV variations, the planet parameters, and the evaluation
of the dynamical stability of the system.
In a forthcoming paper (Damasso et al., in prep.) we will present a 
complete analysis of the XO-2 system.

\section{Observations and data reduction}
\label{s:obsred}

The system XO-2S was observed with HARPS-N at 63 individual epochs between April 2013 and May 2014.
The Th-Ar simultaneous calibration was not used to avoid contamination of the stellar spectrum by the lamp 
lines (which might affect a proper spectral analysis; Sect.~\ref{s:param}).
Nevertheless, the drift correction with respect to the reference calibration shows a dispersion of just 0.8~m~s$^{-1}$,
which is of very limited impact considering the typical photon-noise RV error of 2.2~m~s$^{-1}$.

The reduction of the spectra and the RV measurements were obtained using the latest version (Nov.\ 2013)
of the HARPS-N instrument data reduction software pipeline and the K5 mask. 
The measurement of the RVs is based on the
weighted cross-correlation function (CCF) method \citep{1996A&AS..119..373B,2002A&A...388..632P}.

\section{Stellar parameters}
\label{s:param}

The extracted spectra of XO-2S were coadded to produce a merged spectrum with a peak S/N ratio of about 200 at 550~nm.
We determined the stellar parameters using implementations of both the
equivalenth width and the spectral synthesis methods, as described in \citet{2014A&A...564L..13E} and \citet{2013A&A...556A.150S}.
We also used the infrared flux method to estimate $T_{\rm eff}$. The results agree very well with each other.
The adopted parameters are the weighted averages of the individual results (Table \ref{t:param}).
Full details will be given in Damasso et al.\ (in prep.).

The stellar mass, radius, and age were estimated using the Yonsei-Yale evolutionary tracks \citep{2004ApJS..155..667D} that
match the effective temperature, iron abundance, and surface gravity of XO-2S (Table \ref{t:param}).
The adopted errors include an additional 5\% in mass and 3\%
in radius added in quadrature to the formal errors to take systematic uncertainties in stellar models
into account \citep{2011MNRAS.417.2166S}.

The star was found to be a slow rotator (using a preliminary calibration of the FWHM of the CCF and spectral synthesis) that exhibits low levels of chromospheric emission.
Other age diagnostics (kinematics, lack of detectable amounts of lithium) support the old age
of the system and confirm the findings by \citet{2007ApJ...671.2115B}.

\begin{table}
   \caption[]{XO-2S stellar parameters}
     \label{t:param}
       \centering
       \begin{tabular}{cc}
         \hline
         \noalign{\smallskip}
         Parameter  &  Value   \\
         \noalign{\smallskip}
         \hline
         \noalign{\smallskip}

$T_{\rm eff}$ (K)     & $5399 \pm 55$   \\
$\log g$ (cgs)        & $4.43 \pm 0.08$   \\
${\rm [Fe/H]}$ (dex)  & $0.39 \pm 0.05$   \\
Microturb.  ($\kms$)  & $0.9 \pm 0.1$   \\
Mass  ($\Msun$)       & $0.982 \pm 0.054$       \\
Radius ($\Rsun$)      & $1.02_{-0.07}^{+0.09}$   \\
Age (Gyr)             & $7.1_{-2.9}^{+2.5}$      \\
$v \sin i$ ($\kms$)   & $1.7\pm0.4$ \\
$\log R_{HK}$         & -5.03   \\
          \noalign{\smallskip}
         \hline
      \end{tabular}

\end{table}

\section{RV variations and their origin}
\label{s:rv}

The relative RV time series is shown in Fig.~\ref{f:rv}. We report in Table~\ref{t:rv} the full dataset.
The RVs show a dispersion of 34.3~m~s$^{-1}$, significantly exceeding the measurement errors. A Lomb-Scargle periodogram
yields the most significant power at a period of 120~d.
The RV semi-amplitude resulting from a Keplerian fit is $K=58$~m~s$^{-1}$.
Moreover, the post-fit residuals show additional modulations with a period of 18.3 d\footnote{During
eight observing nights, the spectra were obtained twice per night with a separation of a few hours
to investigate the possibility of aliases. These data corroborated the 18.3~d periodicity.}.
A test based on bootstrap random permutation shows that this periodicity is highly significant, with a false-alarm probability lower than $10^{-4}$,
and a corresponding semi-amplitude $K \sim 20$~m~s$^{-1}$. The period value, however, is fairly typical of
rotational periods of old, solar-type dwarfs. Therefore, we performed specific checks to exclude
rotational modulations and other stellar phenomena as the origin of the RV variations.

\begin{figure*}
   \includegraphics[width=15cm]{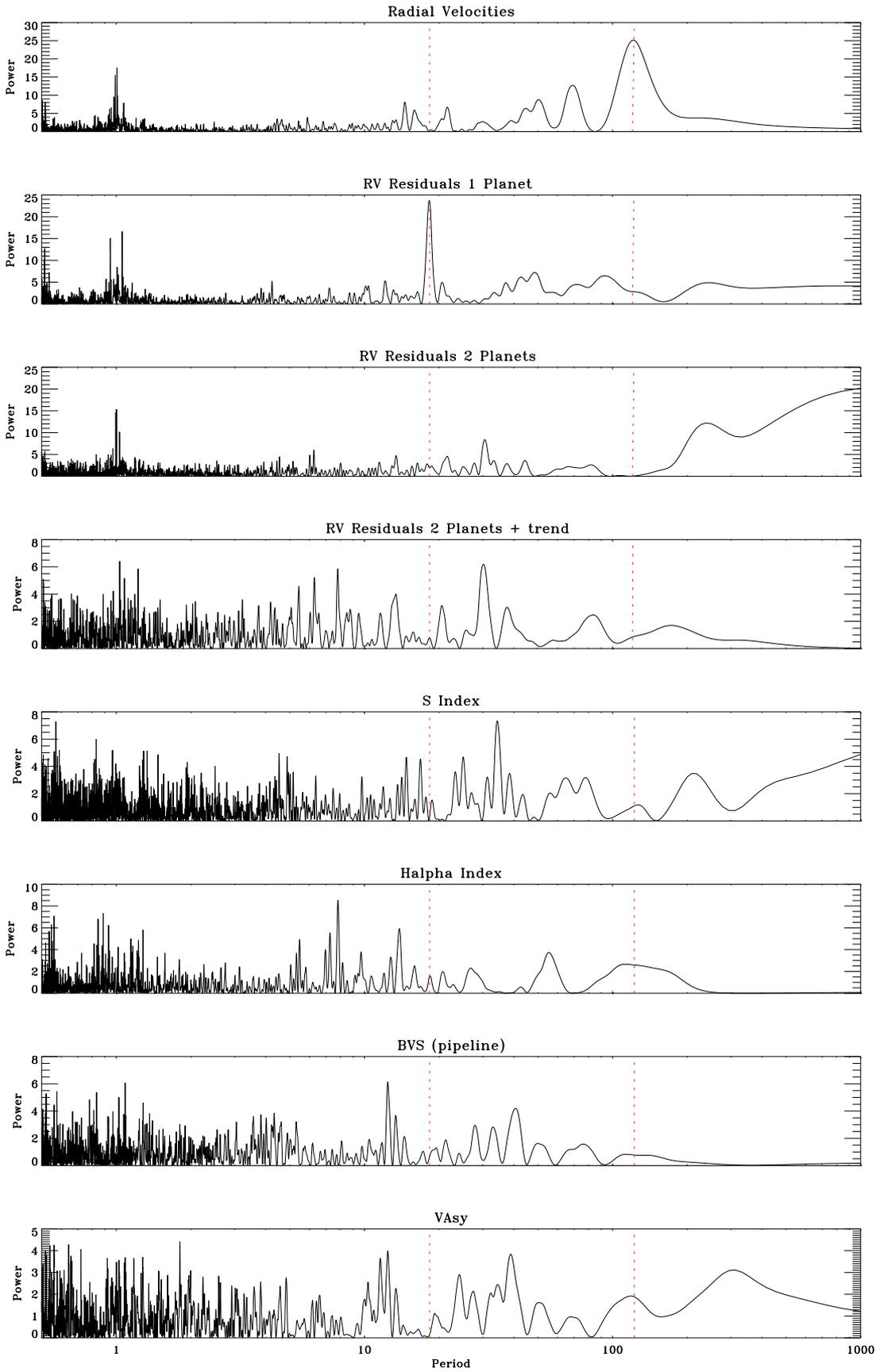}
      \caption{Lomb-Scargle periodogram of RV and activity indicators of \object{XO-2S}. From top to bottom:
               raw RVs, one-planet fit residuals, two-planet fit residuals, two-planet plus linear-trend fit residuals;
               H and K S index, H$\alpha$ index, bisector velocity span, $V_{\rm asy}$ indicator
               \citep[see ][]{2013A&A...557A..93F}. In all panels
               the periodicities of the two planetary companions are marked with vertical dashed lines.}
         \label{f:periodogram}
   \end{figure*}

We first searched for correlations between the RVs and line profile indicators, considering the
bisector velocity span as delivered by the HARPS-N pipeline and our own implementation of the diagnostics
proposed by \citet{2013A&A...557A..93F}.
We also derived indices to measure the chromospheric activity in the CaII H\&K and H${\alpha}$ lines.
None of these indices show either significant power at the RV periods or significant correlations
with the original RVs or the residuals of the one-planet fit (Fig.~\ref{f:periodogram}).

To further exclude stellar variability as the origin of the observed RV signals,
we obtained accurate multi-band photometry at the Asiago, Serra La Nave and Aosta Valley observatories.
The latter was used to monitor XO-2S during 42 nights
from 2 Dec 2013 to 8 Apr 2014 in the {\em I} band, following the observations and data
reduction procedure of the APACHE program \citep{2013EPJWC..4703006S,2013EPJWC..4717001C}.
At Serra La Nave we observed in {\em B, V, R, I} bands on four nights in 2013-2014.
We also analyzed data of the XO-2 system acquired at Asiago
in the context of the TASTE program \citep{2011A&A...527A..85N} during ten nights in which transits of
the hot-Jupiter in front of XO-2N occurred.
The APACHE  time series shows a scatter of 0.003 mag (nightly averages), without significant power
at the period of the RV variations.
The nightly averages of the Asiago and Serra La Nave datasets show a similar scatter
in the magnitude difference between XO-2N and XO-2S,
confirming the low level of photometric variability of the two stars.
The photometric amplitude due to star-spots associated with an activity-induced RV variation with
a semi-amplitude of 20~m~s$^{-1}$ 
\citep[derived using Eq.~1 of ][]{2007A&A...473..983D} is about 3\%, which is
one order of magnitude larger than the observed photometric variability.

In summary, the tests performed allowed us to rule out phenomena related to stellar activity and rotation
as the cause of the observed RV variations with periods of $\sim$120 and $\sim$18~d. We conclude that these are
due to Keplerian motions, which are characterized in Sect.~\ref{s:orbit}.

\section{Orbital parameters and dynamical stability}
\label{s:orbit}

\begin{figure}
   \includegraphics[width=7.5cm]{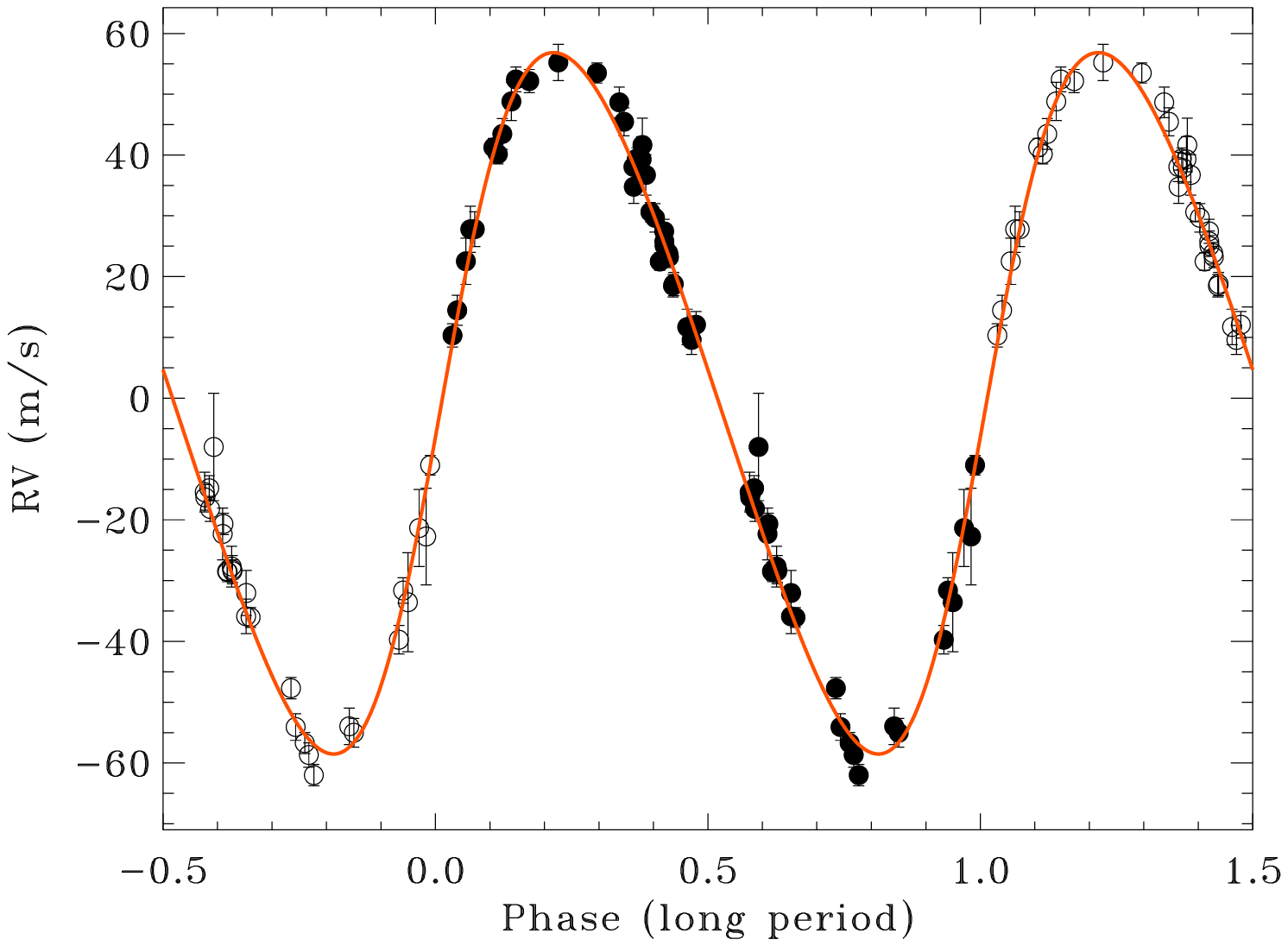}
   \includegraphics[width=7.5cm]{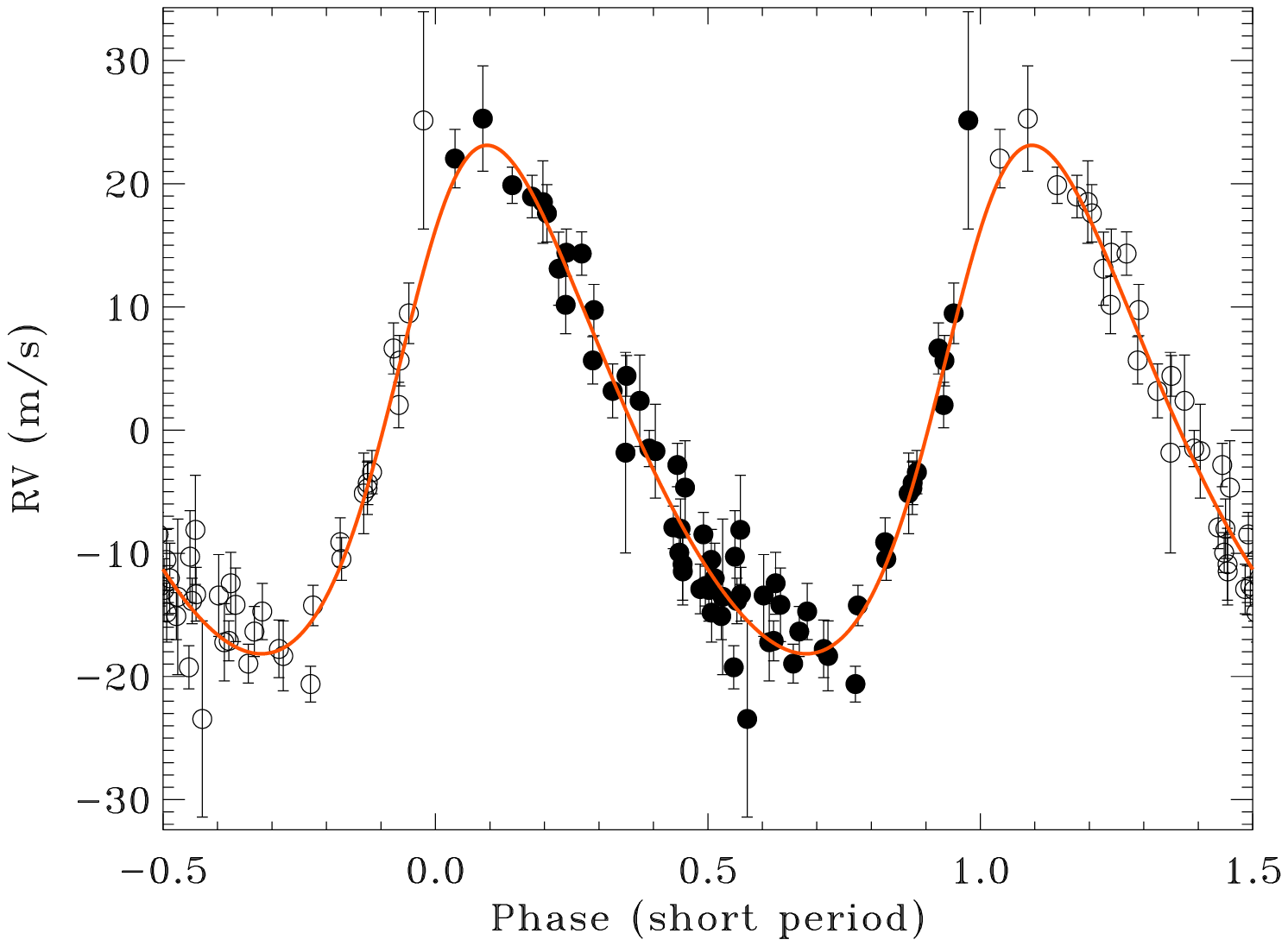}
      \caption{Phased plot for the two planets orbiting XO-2S. In each panel, the RV signal of the other
               object and the long-term trend are subtracted. Upper panel: long-period planet (XO-2Sc).
               Lower panel: short-period planet (XO-2Sb).}
         \label{f:phase}
   \end{figure}

Orbital parameters and associated uncertainties were determined with a Bayesian
differential evolution Markov chain Monte Carlo analysis
of the HARPS-N data \citep{TerBraak2006, 2013PASP..125...83E}
by maximizing a Gaussian likelihood function \citep[e.g.,][]{2006ApJ...642..505F}.
Our two-planet model has 13 free parameters: periastron
epoch $T_0$, orbital period $P$, $K$, $\sqrt{e}\,\cos\omega$, and $\sqrt{e}\,\sin\omega$ of
the two planets XO-2S\,b and \,c, $e$ and $\omega$
being the eccentricity and the argument of periastron,
the systemic velocity $\gamma$, a slope, and a jitter term \citep[e.g.,][]{2005ApJ...631.1198G},
which includes RV scatter that is possibly induced by stellar variability and/or instrumental
noise that exceeds the nominal error bars.
Uninformative priors were used for all parameters.
Twenty-six chains were run simultaneously, and were stopped after
convergence and good mixing of the chains were reached according to
\citet{2006ApJ...642..505F}. Burn-in steps were removed following \citet{2013PASP..125...83E}.
The medians of the parameter posterior distributions
and their $34.13 \%$ intervals are taken as final values  
and their 1-$\sigma$ uncertainties, respectively.
The results are listed in Table \ref{t:fit} and shown in Fig.~\ref{f:phase}.

\begin{figure}
   \includegraphics[width=7.8cm]{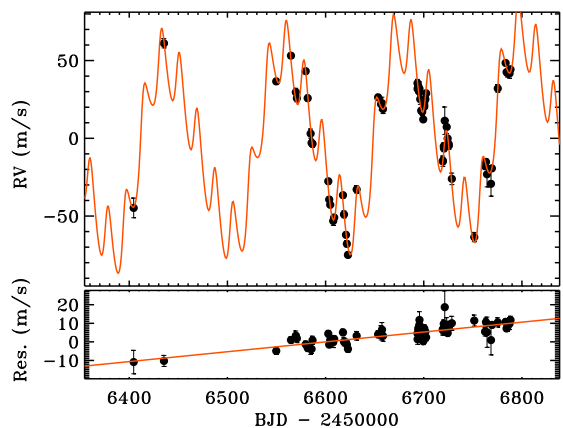}
      \caption{Upper panel: relative RVs of \object{XO-2S}.
               The overplotted line is the best-fit Keplerian solution for two planets and
               a linear trend. Lower panel: residuals of \object{XO-2S} RVs after removing the
               contribution of the two planets XO-2Sb and XO-2Sc.
               The overplotted line is the best-fit linear trend.}
         \label{f:rv}
   \end{figure}

The two planets have moderately eccentric orbits:
$e=0.180 \pm 0.035$ (XO-2S\,b) and
$0.153 \pm 0.010$ (XO-2S\,c).
The inferred minimum masses are $0.259 \pm 0.014$ and $1.370 \pm 0.053~\Mjup$
for planets \,b and \,c, respectively. Interestingly, we found evidence at the 6-$\sigma$ level for a long-term trend
of $0.053 \pm 0.009$~m~ s$^{-1}$\,d$^{-1}$ (Fig.~\ref{f:rv}) in the residuals of the two-planet fit.
This is likely due to an additional companion whose nature remains to be established.
The trend cannot be due to the wide companion XO-2N because of its very large separation (a slope of the
order of $10^{-5}$ ~m~ s$^{-1}$\,d$^{-1}$ is expected).
The orbital parameters of the two planetary companions are only marginally affected by the
inclusion of the long-term RV slope.

\begin{table}
\setcounter{table}{2}
   \caption[]{Orbital parameters}
     \label{t:fit}
       \centering
       \begin{tabular}{ccc}
         \hline
         \noalign{\smallskip}
         Parameter  &  XO-2Sb  &  XO-2Sc  \\
         \noalign{\smallskip}
         \hline
         \noalign{\smallskip}

$P$ (d)                     &   $ 18.157 \pm 0.034        $     &  $ 120.80 \pm 0.34   $    \\
$K$ ($\ms$)                      &   $ 20.64 \pm 0.85 $             &  $ 57.68 \pm 0.69  $        \\
$\sqrt{e} \sin \omega$         &   $ -0.314_{-0.052}^{+0.059}$     &  $ -0.388_{-0.012}^{+0.013}$  \\
$\sqrt{e} \cos \omega$         &   $  0.282_{-0.065}^{+0.054}$     &  $ -0.038_{-0.034}^{+0.033}$  \\
\textit{e}                     &   $ 0.180 \pm 0.035 $             &  $ 0.1528_{-0.0098}^{+0.0094}$   \\
$\omega$ (deg)                 &   $ 311.9 \pm 9.5  $              &  $ 264.5 \pm 4.9 $       \\
$T_0$ ($\rm BJD_{TDB}$-2450000)   &   $ 6413.11_{-0.86}^{+0.82}$      &  $ 6408.1_{-1.9}^{+1.8}$      \\
{\bf $T_c$ ($\rm BJD_{TDB}$-2450000) } & $6419.30 \pm 0.53$         &   $6471.02_{-0.90}^{+0.85}$     \\ 
slope (m s$^{-1}$ d$^{-1}$)                  & \multicolumn{2}{c}{  $0.0531 \pm0.0087$ }   \\
$\gamma$ (km s$^{-1}$)                   & \multicolumn{2}{c}{  $46.543 \pm 0.001$ }  \\
jitter ($\ms$)                 & \multicolumn{2}{c}{  $ 1.80 \pm 0.43$ }  \\
rms  ($\ms$)                   & \multicolumn{2}{c}{  $3.1 $ } \\
$m\sin i$  ($\Mjup$)          &   $0.259 \pm 0.014$               & $1.370 \pm 0.053$ \\
$a$ (AU)                       &  $0.1344 \pm 0.0025$              & $0.4756 \pm 0.0087$  \\
          \noalign{\smallskip}
         \hline
      \end{tabular}
\end{table}

Finally, we performed a dynamical analysis of the proposed best fit of the planetary system. We have numerically integrated
the nominal system (assuming nearly coplanar orbits) over 100 Myr using the symplectic integrator SyMBA
\citep{1998AJ....116.2067D} 
and found a stable quasi-periodic evolution. 
We also used the frequency map analysis, as described in
\citet{2005A&A...442..359M}, to explore the stability properties close to the nominal solution.
We found very low diffusion values when sampling the initial conditions in a random way within the error bars of
all orbital elements and of the estimated planetary and stellar masses.
This indicates that the system is located in a robustly stable area of the phase space.

\section{Discussion and conclusions}
\label{s:discussion}

Based on high-precision HARPS-N Doppler measurements,
we presented the discovery of a planetary system composed of a planet slightly more massive than Jupiter
at 0.48~au and a Saturn-mass planet at 0.13~au around XO-2S, the wide companion of the transiting hot-Jupiter host XO-2N.
The two planetary orbits are moderately eccentric and were found to be dynamically stable.
There are also indications of a long-term trend in the RVs. A longer time baseline of the observations is needed
to infer its origin.
The two planets have periods intermediate between the hot-Jupiter pile-up and the rise in frequency at $a \ge 1$~au, which means that they are located in the so-called period-valley in the distribution of known exoplanets \citep{2010ApJ...722.1854W}. 
The geometric transit probabilities for the inner and outer planets are 3\% and 1\%, respectively.
However, the available photometric data and the uncertainties on the transit epochs  (Table \ref{t:fit})
do not allow us to set any meaningful constraints on the occurrence of their transits.

This is the first confirmed case of a wide binary whose components both host planets, 
one of which is transiting.
The two stars are almost twins
but the planetary systems around them are not, with one component hosting a hot-Jupiter and the other one
hosting a Jupiter and a Saturn-mass planet in wider orbits.
This fact makes this system a special laboratory for our understanding
of planet formation processes and the influence of specific classes of planets on their parent stars, 
for example in terms of alteration of the chemical composition \citep{2010A&A...521A..33R} and angular momentum
\citep{2010A&A...512A..77L,2014A&A...565L...1P}. We plan to address these questions in forthcoming publications.

\begin{acknowledgements}
The GAPS project in Italy acknowledges support from INAF through the ``Progetti Premiali''
funding scheme of the Italian Ministry of Education, University, and Research.
The Aosta Valley Observatory is supported by the Regional Government of
the Aosta Valley, the Town Municipality of Nus and the Mont Emilius
Community. JMC is supported by a grant of the EU-ESF,
the Autonomous Region of the Aosta Valley and the Italian
Ministry of Labour and Social Policy. We thank ASI (through contracts
I/037/08/0 and I/058/10/0) and the Fondazione CRT for their support to the
APACHE Project.
DG acknowledges support from the
EU FP7 under grant agreement n. 267251.
NCS acknowledges support from Funda\c{c}\~ao para a
Ci\^encia e a Tecnologia (FCT, Portugal) through the FEDER funds in the program
COMPETE, as well as through national funds, in the form of grants
reference RECI/FIS-AST/0176/2012 (FCOMP-01-0124-FEDER-027493), and
RECI/FIS-AST/0163/2012 (FCOMP-01-0124-FEDER-027492), and in
the form of the Investigador FCT contract reference IF/00169/2012
and POPH/FSE (EC) by the FEDER funding through the program "Programa
Operacional de Factores de Competitividade - COMPETE.
NCS furthermore acknowledges the support from the ERC/EC
under the FP7 through Starting Grant agreement n.~239953.
We thank the TNG staff for help with the observations.

\end{acknowledgements}

\bibliography{xo2s}
\bibliographystyle{aa}

\begin{table*}
\setcounter{table}{1}
   \caption[]{Radial velocities, bisector velocity span, and $\log R_{HK}$ of XO-2S. The errors on the bisector velocity span
              are about twice as high as the errors on radial velocities.}
     \label{t:rv}
       \centering
       \begin{tabular}{cccccc}
         \hline
         \noalign{\smallskip}
        $ \rm BJD_{UTC} - 2\,450\,000$  &  RV & error  & Bisector & $\log R_{HK}$ & err $\log R_{HK}$ \\
                      &  (km s$^{-1}$) & (km s$^{-1}$) &  (km s$^{-1}$) &         &  \\
         \noalign{\smallskip}
         \hline
         \noalign{\smallskip}

   6404.52027    &  46.4982    &   0.0063    &   0.0069    &   -5.478    &    0.584   \\
   6435.37047    &  46.6041    &   0.0030    &   0.0111    &   -5.263    &    0.111   \\
   6549.73055    &  46.5796    &   0.0019    &   0.0034    &   -5.126    &    0.045   \\
   6564.73154    &  46.5961    &   0.0017    &   0.0078    &   -5.020    &    0.028   \\
   6569.69948    &  46.5727    &   0.0025    &   0.0078    &   -5.054    &    0.057   \\
   6570.74915    &  46.5688    &   0.0023    &   0.0051    &   -5.134    &    0.059   \\
   6579.73878    &  46.5862    &   0.0017    &   0.0097    &   -5.109    &    0.035   \\
   6581.76437    &  46.5688    &   0.0019    &   0.0107    &   -5.006    &    0.034   \\
   6584.75753    &  46.5460    &   0.0029    &   0.0039    &   -5.070    &    0.071   \\
   6585.72086    &  46.5401    &   0.0024    &  -0.0031    &   -5.083    &    0.055   \\
   6586.70194    &  46.5395    &   0.0022    &   0.0002    &   -5.000    &    0.040   \\
   6602.73935    &  46.5154    &   0.0018    &   0.0061    &   -5.053    &    0.033   \\
   6603.67615    &  46.5036    &   0.0016    &   0.0123    &   -4.993    &    0.025   \\
   6604.72411    &  46.5003    &   0.0019    &   0.0075    &   -5.053    &    0.035   \\
   6607.76170    &  46.4898    &   0.0028    &   0.0172    &   -5.100    &    0.075   \\
   6608.75358    &  46.4921    &   0.0016    &   0.0040    &   -5.061    &    0.030   \\
   6617.70822    &  46.5064    &   0.0018    &   0.0050    &   -4.981    &    0.027   \\
   6618.73859    &  46.4941    &   0.0022    &   0.0118    &   -5.053    &    0.045   \\
   6620.76267    &  46.4809    &   0.0017    &   0.0085    &   -5.053    &    0.032   \\
   6621.65931    &  46.4751    &   0.0020    &   0.0055    &   -5.090    &    0.045   \\
   6622.77386    &  46.4680    &   0.0018    &   0.0061    &   -4.996    &    0.029   \\
   6631.64148    &  46.5101    &   0.0024    &   0.0071    &   -5.008    &    0.048   \\
   6653.51105    &  46.5693    &   0.0019    &   0.0055    &   -5.100    &    0.043   \\
   6656.47923    &  46.5649    &   0.0038    &   0.0074    &   -5.004    &    0.091   \\
   6657.46522    &  46.5656    &   0.0038    &   0.0014    &   -5.198    &    0.143   \\
   6658.45326    &  46.5618    &   0.0029    &   0.0104    &   -5.077    &    0.075   \\
   6693.63057    &  46.5785    &   0.0024    &   0.0036    &   -5.006    &    0.049   \\
   6693.70485    &  46.5749    &   0.0027    &   0.0060    &   -4.913    &    0.050   \\
   6694.39223    &  46.5767    &   0.0018    &   0.0095    &   -4.976    &    0.028   \\
   6694.65433    &  46.5742    &   0.0025    &   0.0103    &   -5.000    &    0.048   \\
   6695.44269    &  46.5730    &   0.0037    &   0.0117    &   -4.901    &    0.064   \\
   6695.62098    &  46.5749    &   0.0044    &   0.0085    &   -5.106    &    0.131   \\
   6696.39858    &  46.5681    &   0.0033    &   0.0027    &   -4.963    &    0.064   \\
   6697.38288    &  46.5608    &   0.0016    &   0.0021    &   -5.026    &    0.026   \\
   6698.40230    &  46.5600    &   0.0023    &   0.0011    &   -4.995    &    0.042   \\
   6699.45786    &  46.5551    &   0.0014    &   0.0046    &   -5.038    &    0.023   \\
   6700.44799    &  46.5647    &   0.0020    &   0.0085    &   -4.987    &    0.034   \\
   6700.48597    &  46.5633    &   0.0017    &   0.0136    &   -5.027    &    0.030   \\
   6701.36430    &  46.5677    &   0.0018    &   0.0043    &   -5.009    &    0.032   \\
   6701.50593    &  46.5683    &   0.0018    &   0.0014    &   -5.033    &    0.031   \\
   6702.39608    &  46.5719    &   0.0019    &   0.0005    &   -5.051    &    0.035   \\
   6719.38765    &  46.5283    &   0.0033    &   0.0119    &   -5.039    &    0.099   \\
   6719.51119    &  46.5284    &   0.0021    &   0.0006    &   -4.999    &    0.039   \\
   6720.38049    &  46.5379    &   0.0021    &   0.0040    &   -5.007    &    0.038   \\
   6720.58213    &  46.5365    &   0.0021    &   0.0038    &   -5.064    &    0.043   \\
   6721.37526    &  46.5543    &   0.0088    &   0.0067    &   -5.030    &    0.254   \\
   6723.35649    &  46.5503    &   0.0043    &   0.0035    &   -5.008    &    0.104   \\
   6724.33597    &  46.5429    &   0.0015    &  -0.0024    &   -4.984    &    0.021   \\
   6725.35840    &  46.5394    &   0.0034    &   0.0169    &   -5.137    &    0.101   \\
   6725.49169    &  46.5382    &   0.0023    &   0.0028    &   -5.028    &    0.047   \\
   6728.58121    &  46.5169    &   0.0037    &   0.0116    &   -4.985    &    0.085   \\
   6751.42778    &  46.4795    &   0.0030    &   0.0046    &   -5.008    &    0.068   \\
   6762.42972    &  46.5252    &   0.0023    &  -0.0014    &   -5.053    &    0.056   \\
   6763.36878    &  46.5278    &   0.0021    &   0.0069    &   -4.962    &    0.037   \\
   6764.42477    &  46.5199    &   0.0081    &   0.0130    &   -4.899    &    0.192   \\
   6768.47748    &  46.5137    &   0.0080    &   0.0277    &   -4.836    &    0.197   \\
   6769.36163    &  46.5237    &   0.0016    &   0.0067    &   -5.061    &    0.032   \\
   6775.36013    &  46.5751    &   0.0025    &   0.0072    &   -5.150    &    0.075   \\
   6783.37304    &  46.5915    &   0.0015    &   0.0039    &   -5.033    &    0.026   \\
   6784.37312    &  46.5856    &   0.0016    &   0.0026    &   -5.039    &    0.031   \\
   6785.36182    &  46.5849    &   0.0025    &   0.0037    &   -5.178    &    0.082   \\
   6787.36842    &  46.5847    &   0.0031    &  -0.0020    &   -5.126    &    0.102   \\
   6788.37059    &  46.5873    &   0.0020    &   0.0041    &   -5.009    &    0.042   \\

          \noalign{\smallskip}
         \hline
      \end{tabular}

\end{table*}

\end{document}